\documentclass{aa}  
\usepackage{graphicx}
\usepackage{txfonts}
\usepackage[colorlinks,citecolor=blue,urlcolor=blue,filecolor=blue,linkcolor=blue]{hyperref}
\usepackage{amsmath}
\usepackage{amssymb}
\usepackage{upgreek}
\usepackage{natbib}
\usepackage{xcolor}
\usepackage{tikz}
\usepackage{float}
\usepackage{algorithm2e}

\def\Ddedisp{\textsc{dedisp}}
\def\Dtdd{\textsc{tdd}}
\def\Dfdd{\textsc{fdd}}

\begin{document} 
   \title{Fourier-domain dedispersion}

   \author{C.\ G.\ Bassa 
     \and
     J.\ W.\ Romein 
     \and
     B.\ Veenboer 
     \and
     S. van der Vlugt 
     \and
     S.\ J.\ Wijnholds 
   }

   \institute{ASTRON, Netherlands Institute for Radio Astronomy, Oude
     Hoogeveensedijk 4, 7991 PD, Dwingeloo, The Netherlands }

   \date{Received \today; Accepted \today}

   \abstract{We present and implement the concept of the
     Fourier-domain dedispersion (FDD) algorithm, a brute-force
     incoherent dedispersion algorithm. This algorithm corrects the
     frequency-dependent dispersion delays in the arrival time of
     radio emission from sources such as radio pulsars and fast radio
     bursts. Where traditional time-domain dedispersion algorithms
     correct time delays using time shifts, the FDD algorithm performs
     these shifts by applying phase rotations to the
     Fourier-transformed time-series data. Incoherent dedispersion to
     many trial dispersion measures (DMs) is compute, memory-bandwidth
     and I/O intensive and dedispersion algorithms have been
     implemented on Graphics Processing Units (GPUs) to achieve high
     computational performance. However, time-domain dedispersion
     algorithms have low arithmetic intensity and are therefore often
     memory-bandwidth limited. The FDD algorithm avoids this
     limitation and is compute limited, providing a path to exploit
     the potential of current and upcoming generations of GPUs. We
     implement the FDD algorithm as an extension of the
     \textsc{dedisp} time-domain dedispersion software. We compare the
     performance and energy-to-completion of the FDD implementation
     using an NVIDIA Titan RTX GPU against the standard as well as an
     optimized version of \textsc{dedisp}. The optimized
     implementation already provides a factor of 1.5 to 2 speedup at
     only 66\% of the energy utilization compared to the original
     algorithm. We find that the FDD algorithm outperforms the
     optimized time-domain dedispersion algorithm by another 20\% in
     performance and 5\% in energy-to-completion when a large number
     of DMs ($\gtrsim 512$) are required. The FDD algorithm provides
     additional performance improvements for FFT-based periodicity
     surveys of radio pulsars, as the Fourier transform back to the
     time domain can be omitted. We expect that this computational
     performance gain will further improve in the future since the
     Fourier-domain dedispersion algorithm better matches the trends
     in technological advancements of GPU development.}
   

   \maketitle

   \section{Introduction}\label{sec:1}
   Radio waves propagating through the ionized intergalactic,
   interstellar and/or interplanetary medium undergo dispersion, which
   introduces a frequency-dependent time delay between emission at the
   source and reception at the telescope \citep[e.g.][]{lk12}. Hence,
   pulsed radio signals from astrophysical sources like radio pulsars
   and fast radio bursts \citep[see][for reviews]{kk16,phl19} have a
   distinctive dispersion measure (DM), directly related to the
   electron column density between the source and the observer. If not
   properly corrected for, the dispersion delays lead to smearing and
   hence loss of signal-to-noise of the received radio pulses.

   Correcting for the dispersion is called dedispersion, and can be
   performed using two techniques, depending on the state of the radio
   observation. For Nyquist-sampled time-series data of orthogonal
   polarizations (voltage data hereafter), the phase information can
   be used to \textit{coherently dedisperse} the signals through a
   convolution with the inverse of the response function of dispersion
   \citep{han71,hr75}. This technique completely removes the effects
   of dispersion. With coherent dedispersion, dedispersion is
   performed before squaring the orthogonal polarizations signals to
   form the Stokes parameters. For signals that have already been
   squared to form the Stokes parameters, the \textit{incoherent
     dedispersion} method is used. The time-series data of finite
   channels are shifted in time to correct for the dispersion delay
   between channels, while dispersive smearing within channels remains
   uncorrected. Hence, the size of frequency channels and time samples
   is typically optimized to minimize dispersive smearing within
   channels.

   In surveys for new radio pulsars or fast radio bursts, the
   dispersion measure of a new source is a priori unknown and the
   input data need to be dedispersed to many, usually thousands, of
   trial dispersion measures. The majority of current surveys perform
   incoherent dedispersion on Stokes I time-series, typically with
   sample times between 50 to 100\,$\upmu$s, and channel sizes of 0.1
   to 0.5\,MHz (\citealt{kjs+10,slr+14,lbh+15,bcm+16}, see also
   \citealt{lbc+16}), as this requires lower data rates and less
   storage compared to voltage data. Only at the lowest observing
   frequencies ($\nu\lesssim300$\,MHz), voltage data are recorded to
   perform a combination of coherent and incoherent dedispersion
   \citep{bph17,bph+17,pbh+17}.

   The computational cost to incoherently dedisperse is significant, as
   the input data typically consist of $N_\nu\sim10^3$ frequency
   channels for $N_t\sim10^6$ time samples, and needs to be
   dedispersed to $N_\mathrm{DM}\sim10^4$ dispersion measures. Hence,
   algorithms have been developed to either reduce the computational
   cost by reducing the complexity, by using e.g.\ a tree algorithm
   \citep{tay74} or a piece-wise linear approximation of the dispersion
   delay \citep{ran01}, a transformation algorithm \citep{zo17} and/or
   by using hardware acceleration of the direct, brute force,
   dedispersion approach. A review of these algorithms, and an
   implementation of the direct dedispersion algorithm on graphics
   processing units (GPUs), is presented by \citet{Barsdell_2012}.

   \citet{Sclocco_2016} found that the direct, brute force,
   dedispersion algorithm on many-core accelerators is inherently
   memory-bandwidth bound, due to the low arithmetic intensity of the
   required calculations. Hence, the direct dedispersion algorithm
   does not optimally benefit from advances in the computing
   performance of many-core accelerators. In this paper we investigate
   the feasibility of the direct dedispersion approach through, what
   we call, \textit{Fourier-domain dedispersion}, where the time
   delays due to dispersion are performed as phase rotations in the
   Fourier transform of the time-series data of each
   channel\footnote{To avoid confusion with observing frequencies, we
   use the term Fourier-domain instead of frequency-domain.}.

   We present the description of the Fourier-domain dedispersion
   algorithm in \S\,\ref{sec:2} and compare it to the time-domain
   variant. \S\,\ref{sec:3} provides a GPU implementation of the
   Fourier-domain dedispersion algorithm along with an optimization of
   the GPU accelerated time-domain direct dedispersion algorithm by
   \citet{Barsdell_2012}, the results for the different
   implementations are compared in \S\,\ref{sec:4}. We discuss and
   conclude our feasibility study in \S\,\ref{sec:5}.

   \section{Algorithm description}\label{sec:2}
   \subsection{Time-domain dedispersion}
   Dispersion in the ionized intergalactic, interstellar and
   interplanetary medium leads to delays in the arrival time of radio
   signals, The magnitude of this delay depends on the frequency of
   the radio signal observed. For a given dispersion measure
   $\mathrm{DM} = \int n_e \mathrm{d}l$, the path integral over the
   electron density $n_e$ along the line of sight, signals arrive with
   a time delay
   \begin{equation}
     \Delta t \left ( \nu, \mathrm{DM} \right ) = \mathrm{DM} ~
     k_\mathrm{DM} \left ( \nu^{-2} - \nu_0^{-2} \right ),
     \label{eq:dispersion_delay}
   \end{equation}
   where $\nu$ is the observing frequency and $\nu_0$ is the reference
   frequency. The proportionality constant is $k_\mathrm{DM} = (2.41
   \times 10^{-4})^{-1}$\,MHz$^2$\,cm$^3$\,pc$^{-1}$\,s
   \citep{mt72,kul20}.

   The standard incoherent dedispersion method takes an input dynamic
   spectrum $I \left ( t, \nu \right)$ with Stokes-I values as a
   function of observing time $t$ and observing frequency $\nu$ and
   shifts the time series at the individual observing frequencies to
   the nearest time sample given the delays described in
   Eqn.\,\ref{eq:dispersion_delay} based on an assumed DM. This
   results in a new dynamic spectrum in which all time series are
   aligned for the assumed DM. The shifted time series are then summed
   over observing frequency to give the intensity as function of time
   observed at the assumed DM
   \begin{equation}
     I\left ( t, \mathrm{DM} \right ) = \sum_\nu^{N_\nu} I \left ( t - \Delta t \left ( \nu, \mathrm{DM} \right ), \nu \right ).
   \end{equation}
   When the DM of the source is not known, this procedure needs to be
   repeated for many trial DMs.

   The brute-force time-domain dedispersion algorithm requires
   significant memory bandwidth while requiring only minimal
   computations once the data are properly aligned. This led us to
   explore the benefits of analyzing the Fourier transform of the time
   series, where the time shifts can be replaced by phase rotations
   through the Fourier shift theorem \citep[e.g.][]{Bracewell_1986},
   resulting in dense matrix multiplications, which can be computed
   efficiently. We describe this Fourier-domain dedispersion (FDD)
   method in the next section.

   \begin{figure*}
     \centering
     \includegraphics[width=\textwidth]{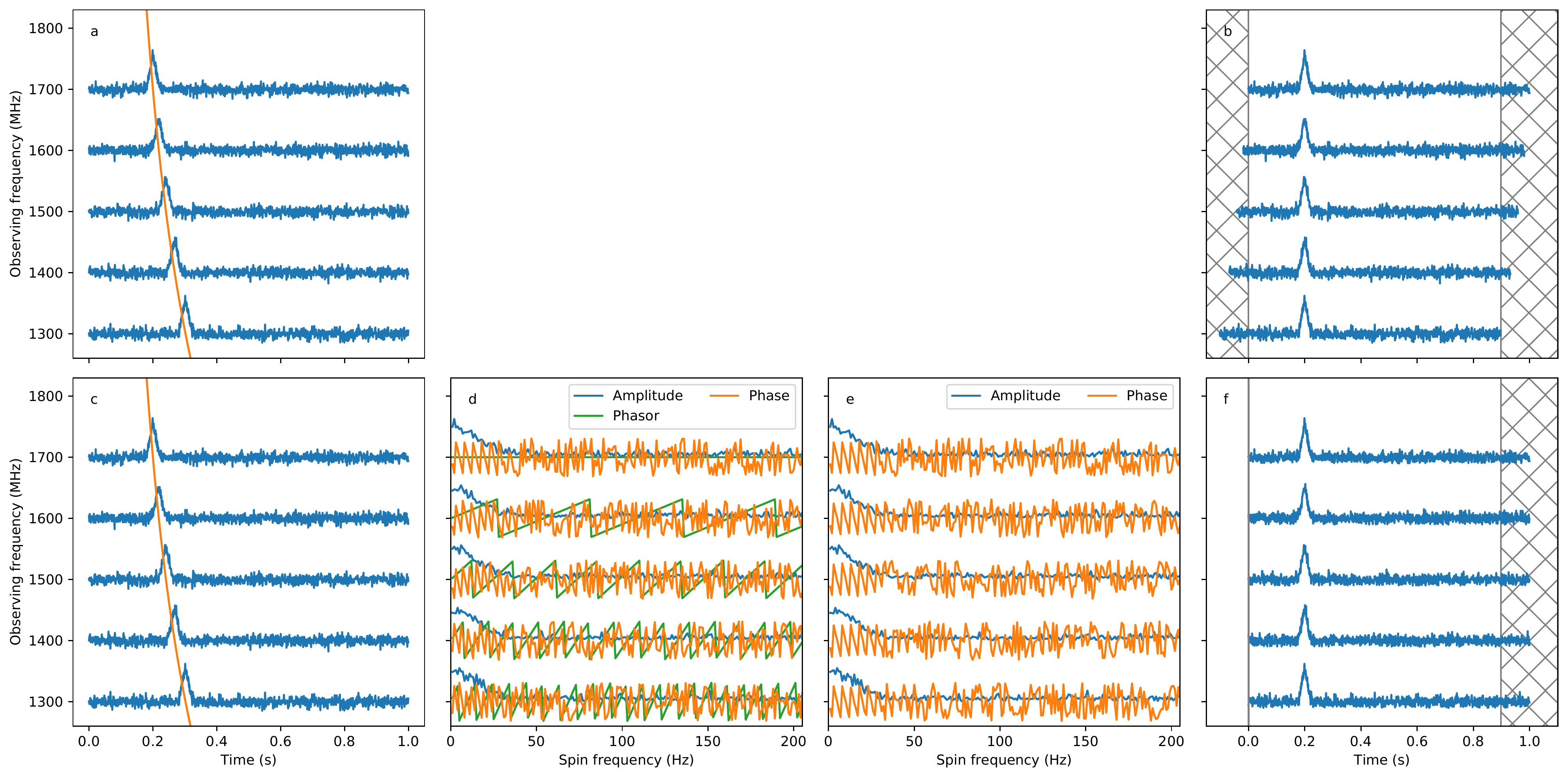}
     \caption{Comparison between time-domain dedispersion (top row)
       and FDD (bottom row) with (a) and (c) representing the input
       dynamic spectrum, (b) showing the dynamic spectrum after
       applying the appropriate delays for each frequency, (d) showing
       the Fourier transform of the time series (amplitude and phase)
       along with the phasors applied in $\mathcal{T}_{\nu \rightarrow
         \mathrm{DM}}$, (e) showing the result after applying the
       phasors and (f) showing the dynamic spectrum after the inverse
       Fourier transform $\mathcal{F}_{t \rightarrow
         f_\mathrm{s}}^{-1}$. The hashed regions in (b) and (f)
       represent data that should be
       discarded.\label{fig:FDD_example}}
   \end{figure*}

   \subsection{Fourier-domain dedispersion}
   \label{ssec:fdd_math}
   In FDD, the time series for each observing frequency is Fourier
   transformed to the Fourier domain of time, frequency. To
   distinguish these frequencies from the observing frequencies of the
   data $\nu$, we will refer to the Fourier transform of the time
   series as the spin frequencies $f_\mathrm{s}$ that would be
   associated with a periodic signal from a radio pulsar. The Fourier
   transform results in a new dynamic spectrum $I \left (
   f_\mathrm{s}, \nu \right )$, in which the data are represented as a
   function of spin frequency $f_\mathrm{s}$ and observing frequency
   $\nu$. We describe this as
   \begin{equation}
     I \left (f_\mathrm{s}, \nu \right ) = \int I \left ( t, \nu \right )
     e^{-2\pi \mathrm{i} f_\mathrm{s} t} \mathrm{d} t = \mathcal{F}_{t
       \rightarrow f_\mathrm{s}} \left \{ I \left (t, \nu \right ) \right
     \},
     \label{eq:mapping_to_fs}
   \end{equation}
   where the Fourier transform operator $\mathcal{F}$ has subscript $t
   \rightarrow f_\mathrm{s}$ to emphasize that this is a mapping from
   time domain to spin-frequency domain instead of the regular 2-D
   Fourier transform. This notation also allows us to use the Fourier
   transform for other mappings later on without causing confusion.

   In (spin) frequency space, the time shifts from
   Eqn.\,\ref{eq:dispersion_delay} can be represented by complex phase
   rotations, which we will refer to as phasors, of the form
   \begin{equation}
     W \left (f_\mathrm{s}, \nu, \mathrm{DM} \right ) = e^{-2 \pi
       \mathrm{i} f_\mathrm{s} \Delta t (\nu, \mathrm{DM} )}.
     \label{eq:phasors}
   \end{equation}
   The dedispersed time-series can now be recovered through the inverse
   Fourier transform $\mathcal{F}_{t \rightarrow f_\mathrm{s}}^{-1}$
   and a summation over observing frequency
   \begin{equation}
     I \left ( t, \mathrm{DM} \right ) = \sum_\nu^{N_\nu} \mathcal{F}_{t \rightarrow
       f_\mathrm{s}}^{-1} \left \{ W \left
     (f_\mathrm{s}, \nu, \mathrm{DM} \right ) I \left ( f_\mathrm{s}, \nu
     \right ) \right \} 
     \label{eq:mapping_to_tDMa}
   \end{equation}
   Due to the linearity of the Fourier transform, the summation and
   inverse Fourier transform can be interchanged, yielding
   \begin{equation}
     I \left ( t, \mathrm{DM} \right ) = \mathcal{F}_{t \rightarrow
       f_\mathrm{s}}^{-1} \left \{ \sum_\nu^{N_\nu} W \left
     (f_\mathrm{s}, \nu, \mathrm{DM} \right ) I \left ( f_\mathrm{s}, \nu
     \right ) \right \}, 
     \label{eq:mapping_to_tDM}
   \end{equation}
   which reduces the number of inverse Fourier transforms from
   $N_\mathrm{DM}\times N_\mathrm{\nu}$ to $N_\mathrm{DM}$.

   Furthermore, it is instructive to define
   \begin{equation}
     I \left ( f_\mathrm{s}, \mathrm{DM} \right ) = \sum_\nu^{N_\nu} W
     \left (f_\mathrm{s}, \nu, \mathrm{DM} \right ) I \left (
     f_\mathrm{s}, \nu \right ) = \mathcal{T}_{\nu \rightarrow
       \mathrm{DM}} \left \{ I \left ( f_\mathrm{s}, \nu \right )
     \right \},
     \label{eq:mapping_to_DM}
   \end{equation}
   since phase rotations provide a mapping from observing frequencies
   to dispersion measures. As the phase rotations depend on spin
   frequency, the data for each spin frequency need to be processed
   independently. For a given spin frequency, this mapping resembles a
   direct Fourier transform. However, it cannot be implemented as a
   fast Fourier transform due to the non-linear dependence of $\Delta
   t$ on observing frequency. Despite these two issues that complicate
   the implementation, we note that Eqn.\,\ref{eq:mapping_to_DM} can
   be viewed as a transformation that maps the space of observing
   frequencies to the space of DM values and denote this
   transformation by $\mathcal{T}_{\nu \rightarrow \mathrm{DM}}$.

   Figure\,\ref{fig:FDD_example} shows a cartoon of the traditional
   time-domain dedispersion algorithm and the Fourier-domain
   algorithm. The panels in the top row show the traditional algorithm
   where time-series at individual observing frequencies are shifted in
   time for a given DM. The steps presented in
   Eqns.\,\ref{eq:mapping_to_fs}, \ref{eq:mapping_to_tDM} and
   \ref{eq:mapping_to_DM} are illustrated in the bottom row of
   Fig.~\ref{fig:FDD_example}. After Fourier transforming the input
   time series for each frequency, shown in panel
   \ref{fig:FDD_example}(c), a spectrum as function of spin frequency
   is obtained, shown in panel \ref{fig:FDD_example}(d), showing the
   amplitude and phase of the complex values. The phase of the Fourier
   transformed time series encodes the arrival time of the pulse. The
   phase slope of the $W \left (f_\mathrm{s}, \nu, \mathrm{DM} \right
   )$ phasor is shown by the green curves in panel
   \ref{fig:FDD_example}(d), and depends on the dispersion delay as
   given in Eqn.\,\ref{eq:phasors}. The transformation
   $\mathcal{T}_{\nu \rightarrow \mathrm{DM}}$ corrects the measured
   phases with the appropriate phase slope. In our example, this
   results in the phases shown in panel \ref{fig:FDD_example}(e). As
   this is a pure phase rotation, the amplitudes are unaltered. After
   applying the inverse Fourier transform to each of the spectra, we
   obtain a shifted time series at each frequency. When the correct
   phasors are applied, this results in the peaks being aligned as
   shown in \ref{fig:FDD_example}(f) and we obtain the same results as
   with time-domain dedispersion as shown in \ref{fig:FDD_example}(b)
   for comparison.

   The use of the Fourier shift theorem and the cyclic nature of the
   Fourier transform contaminates the end of each time series with
   samples from the start. Hence, given a maximum $\mathrm{DM}$
   ($\mathrm{DM}_\mathrm{max}$) and the extrema of the observing band
   ($\nu_\mathrm{min}$ and $\nu_\mathrm{max}$) at least $\Delta
   t_\mathrm{max}=\mathrm{DM}_\mathrm{max} ~ k_\mathrm{DM} \left (
   \nu_\mathrm{min}^{-2} - \nu_\mathrm{max}^{-2} \right )$ at the end
   of each time series should be discarded when dedispersing against
   the highest observing frequency channel, see
   Fig.\,\ref{fig:FDD_example}(f). In the time-domain dedispersion
   algorithm the same amount of data should be discarded, see
   Fig.\,\ref{fig:FDD_example}(b).

   As dispersive smearing increases with increasing $\mathrm{DM}$,
   time-domain dedispersion algorithms down-sample the input time-series
   when the smearing exceeds one or more native time samples. In the
   Fourier-domain algorithm, this down-sampling can be efficiently
   implemented by reducing the length of the inverse Fourier transform
   $\mathcal{F}_{t \rightarrow f_\mathrm{s}}^{-1}$.

   \subsection{Algorithm optimization}
   \label{ssec:alg_optimisation}
   The transformation $\mathcal{T}_{\nu \rightarrow \mathrm{DM}}$
   poses two implementation challenges: the phasors described by
   \eqref{eq:phasors} depend on the spin frequency and the non-linear
   dependence of $\Delta t$ on observing frequency precludes the use
   of the fast Fourier transform. In this section, we present three
   approximations to reduce the computational burden incurred by these
   two challenges.

   Calculation of the phasors for each trial DM, observing frequency
   and spin frequency consumes a significant amount of computing
   resources. Also, an actual implementation of
   \eqref{eq:mapping_to_DM} leads to a matrix-vector product of a
   matrix of phasors for constant $f_\mathrm{s}$ extracted from the
   phasor tensor $W(f_\mathrm{s}, \nu, \mathrm{DM})$ and a vector with
   data values for constant $f_\mathrm{s}$ extracted from the data
   matrix $I(f_\mathrm{s}, \mathrm{DM})$. As GPUs can perform matrix
   multiplications efficiently, it is expected to be beneficial to
   process data for multiple spin frequencies
   simultaneously. Therefore, it may be worthwhile to accept a
   spin-frequency-dependent set of DM values and re-sample those
   afterwards to the desired set of trial DMs. To this end, consider
   the following spin-frequency-dependent DM
   \begin{equation}
     \mathrm{DM}_{f_\mathrm{s}} = \mathrm{DM}_{f_{s,0}} \frac{f_{s,0}}{f_\mathrm{s}}.
     \label{eq:scaled_DMs}
   \end{equation}
   Substitution in \eqref{eq:phasors} gives
   \begin{eqnarray}
     W \left ( f_\mathrm{s}, \nu, \mathrm{DM}_{f_\mathrm{s}} \right )
     & = & \exp \left \{ -2 \pi \mathrm{i} f_\mathrm{s}
     \mathrm{DM}_{f_\mathrm{s}} k_\mathrm{DM} \left ( \frac{1}{\nu^2}
     - \frac{1}{\nu_0^2} \right) \right \} \nonumber\\ & = & \exp
     \left \{ -2 \pi \mathrm{i} f_\mathrm{s} \mathrm{DM}_{f_{s, 0}}
     \frac{f_{s,0}}{f_\mathrm{s}} k_\mathrm{DM} \left (
     \frac{1}{\nu^2} - \frac{1}{\nu_0^2} \right) \right \} \nonumber
     \\ & = & \exp \left \{ -2 \pi \mathrm{i} f_{s,0}
     \mathrm{DM}_{f_{s, 0}} k_\mathrm{DM} \left ( \frac{1}{\nu^2} -
     \frac{1}{\nu_0^2} \right ) \right \} \nonumber \\ & = & \exp
     \left \{ -2 \pi \mathrm{i} f_{s,0} \Delta t ( \nu,
     \mathrm{DM}_{f_{s,0}} ) \right \}.
   \end{eqnarray}
   Note that these phasors are independent of spin frequency, implying
   that these phasors can be applied over a range of spin frequencies
   reasonably close to the reference spin frequency $f_{s,0}$. The
   extent of this range is limited by the tolerable amount of
   contraction or expansion of the set of DM values for spin
   frequencies other than $f_{s,0}$ as given by \eqref{eq:scaled_DMs}.

   Although the structure of the phasors describing the mapping to DM
   space, $\mathcal{T}_{\nu \rightarrow \mathrm{DM}}$, appears to
   resemble the Fourier transform very closely, the exact
   transformation from observing frequencies to DM values cannot be
   represented by a discrete Fourier transform (DFT) matrix, let alone
   a fast Fourier transform (FFT) due to the non-linear dependence of
   the phases of the phasors on observing frequencies. This can be
   remedied by using a first order Taylor approximation for the
   frequency dependence, i.e.,
   \begin{equation}
     \left ( \frac{1}{\nu^2} - \frac{1}{\nu_0^2} \right ) \Bigg | _{\nu_0} \approx \frac{2}{\nu_0^2} - \frac{2 \nu}{\nu_0^3}.
   \end{equation}
   Substitution of this approximation in \eqref{eq:phasors} gives
   \begin{equation}
     W \left ( f_\mathrm{s}, \nu, \mathrm{DM} \right ) = \exp \left \{
     -2 \pi \mathrm{i} f_\mathrm{s} \mathrm{DM} k_\mathrm{DM} \cdot 2
     \left ( \frac{\nu_0 - \nu}{\nu_0^3} \right ) \right \}.
   \end{equation}
   This approximation can be exploited to implement $\mathcal{T}_{\nu
     \rightarrow \mathrm{DM}}$ by a FFT per spin frequency or even
   multiple spin frequencies when combined with the optimization from
   the previous subsection. The observing frequency range over which
   this gives acceptable results depends on DM and spin frequency.

   To be able to treat the $\mathcal{T}_{\nu \rightarrow \mathrm{DM}}$
   transformation as a DFT, and be able to use the advantages of an
   FFT for precision and performance over a DFT, the observing
   frequencies $\nu$ would have to be remapped to a linear scale
   following $\nu^{-2}$. Such a remapping has been used by
   \citet{mlc+01}. This mapping could be obtained through either
   interpolation or inserting empty channels in the linearized scale
   at $\nu^{-2}$ values that do not map to integer observing
   channels. It is likely that such a remapping would increase GPU
   memory usage, as well as introduce some smearing in the dedispersed
   time series. The analysis of this optimization and assessment of its
   performance and accuracy over the direct computation of the
   transform goes beyond the scope of this paper.

   \section{Implementation}\label{sec:3}
   We have implemented Fourier-domain dedispersion as an extension of
   the \textsc{dedisp}
   library\footnote{\url{{https://github.com/ajameson/dedisp}}} by
   \citet{Barsdell_2012}. This library implements the brute-force,
   \textit{direct} incoherent dedispersion algorithm on NVIDIA GPUs
   and is widely used (e.g.\,\citealt{Heimdall, Sclocco_2015}). To
   make a fair comparison, we have updated the \textsc{dedisp} library
   by \citet{Barsdell_2012} to make use of newer hardware and software
   features. In the following, we will refer to the original
   implementation as \Ddedisp~and to the updated implementation as
   \Dtdd~(time-domain dedispersion).

   We refer to our implementation of Fourier-domain dedispersion as
   \Dfdd. For this work we have implemented the algorithm as described
   under \S\,\ref{ssec:fdd_math}, additional optimizations as proposed
   in \S\,\ref{ssec:alg_optimisation}, as well as the use of new GPU
   tensor-core technology are left as future work.

   \citet{Sclocco_2016} found that brute-force dedispersion in the
   time domain has a low arithmetic intensity and is inherently
   memory-bandwidth bound. They show that auto-tuning of kernel
   parameters significantly improves performance on various
   accelerators. For this work we explore the feasibility of
   dedispersion in the Fourier domain and compare performance for a
   representative parameter space according to the use cases outlined
   in the introduction. The performance of time-domain dedispersion is
   predominantly determined by the amount of data reuse and memory
   bandwidth.

   The work of \citet{Barsdell_2012} already shows that major
   performance improvements are achieved when moving the
   implementation from a central processing unit (CPU) to a
   GPU. Performance improvements are mainly attributed to the higher
   memory bandwidth of the GPU. We compare \Ddedisp, \Dtdd~and
   \Dfdd~on the NVIDIA Titan RTX GPU. The full measurement set-up is
   listed in Table \ref{tab:setup}. We refer, in NVIDIA CUDA
   terminology, to the CPU as \textit{host} and to the GPU as
   \textit{device}. The main application runs on the host and launches
   computational kernels on the device. Input and output data are
   copied between the host and device over a PCIe interface.

   Dedispersion is usually part of a pipeline running on a GPU where
   initialization of and input/output to the GPU kernels are part of
   the pipeline. The input to dedispersion is a matrix of size $N_\nu
   \times N_t$ of packed 8-bit data, with $N_\nu$ frequency channels
   and $N_t$ time samples. For time-domain dedispersion these data
   have to be transposed, unpacked and dedispersed. The resulting
   output is a matrix of $N_\mathrm{DM} \times N_t$ 32-bit floating
   point data, with $N_{DM}$ trial-DMs.

   For time-domain dedispersion the input data $I \left ( t, \nu
   \right)$ can be segmented in the time dimension, taking a segment
   of $M_t (< N_t)$ of time samples, for all frequency
   channels. \Ddedisp~implements the above described pipeline with
   batch processing where a segment of input data is copied from host
   to device, the GPU applies the kernel pipelines, parallelizing over
   DMs, and the resulting output data $I \left ( t, \mathrm{DM}
   \right)$ for the segment is copied back to the host before copying
   a new segment of data to the device.

   We extended the \textsc{dedisp} framework with an optional
   \Dtdd~implementation that contains several improvements over
   \Ddedisp. First, the transfers of input and output data were
   overlapped with computations on the GPU, moving the copying of
   input and output out of the critical path. Transfer speeds were
   increased by a factor of 2 to 3 by changing the memory transfers
   from paged to pinned memory. Furthermore, the unpacking and
   transpose operations were combined into one kernel, thus requiring
   only a single pass over the data. Finally, the use of texture
   memory is optionally disabled, as this optimization for old GPUs
   reduces performance on recent GPUs.  These improvements are
   currently available as a
   fork\footnote{\url{https://github.com/svlugt/dedisp}} of the
   \textsc{dedisp} library by \citet{Barsdell_2012} and are being
   integrated with the original library.

   For the implementation of \Dfdd, the input data cannot
   straightforwardly be segmented in time. It follows from
   Eqn.\,\ref{eq:mapping_to_fs} that splitting in the time dimension
   affects the outcome of the FFT. Instead we split input data along
   the frequency-channel dimension (shown as channel job on line
   \ref{alg:fdd-channel} of Algorithm listing~\ref{alg:fdd}). Such an
   input data segment is copied to the GPU, where the data are
   transposed and unpacked, like in the preprocessing step in
   \Dtdd. Next, the data are zero padded such that the number of
   samples becomes a power of two. The last step of preprocessing is
   the Fourier transformation (Eqn.~\ref{eq:mapping_to_fs}) of the
   input samples. Dedispersion is performed as phase rotations
   (Eqn.~\ref{eq:mapping_to_DM}) in the Fourier domain, followed by a
   summation over frequency channels.  The result is transformed back
   to the time domain (Eqn.~\ref{eq:mapping_to_tDM}) before it is
   scaled and transferred back from device to host. Both forward and
   backward Fourier transformations are implemented with the NVIDIA
   \textit{cuFFT} library.  Continuing on Algorithm listing
   \ref{alg:fdd}, we show that within batches of channels the work is
   split into batches of DMs (DM job starting at line
   \ref{alg:fdd-dm_inner1}) to efficiently parallelize the workload on
   the GPU. The backward FFT is implemented as a batched series of 1D
   FFTs for each DM. To allow for efficient batching of the backwards
   FFTs, the output of the dedispersion kernel is buffered on the GPU
   for all batches of channels before it is transformed back to the
   time domain (starting at line \ref{alg:fdd-dm_inner2}).  Due to GPU
   memory size constraints, we cannot process all the DMs at once for
   a large number of DMs. We solve this by processing the DMs in
   batches with an additional $DM\_outer$ dimension (line
   \ref{alg:fdd-dm_outer}). This approach has the advantage that the
   output data are only copied once, after all input for the current
   batch of DMs is processed and postprocessing has taken place. The
   input data, on the other hand, has to be copied and preprocessed
   multiple times for different batches of DMs. We minimize the cost
   of these repeated operations by overlapping them with the copy of
   output data.  Batching of channels and DMs also allows for data
   reuse in the other GPU kernels.\\

   \newfloat{algorithm2e}{t}{p}
   \SetAlgoCaptionLayout{small}
   \SetAlgoCaptionSeparator{:}
   \SetAlCapSty{}
   \SetAlCapFnt{\footnotesize}
   \SetAlCapNameSty{}
   \SetAlCapNameFnt{\footnotesize}
   \SetAlCapSkip{1ex}
   \SetAlgoLined
   \LinesNumbered
   \SetInd{0.4em}{0.4em}
   \newcommand{\cfloat}{complex$<$float$>$~}
   \newcommand{\fft}{f\mkern-3mu f\mkern-2mu t}
   \newcommand{\buffers}{bu\mkern-2mu f\mkern-3mu f\mkern-2mu ers}
   \SetKwProg{Fn}{void}{:}{}
   \SetKwFor{For}{for}{:}{}
   \SetKw{Pragma}{\#pragma}
   
   \definecolor{cBlue}{HTML}{2171B5}
   \definecolor{cOrange}{HTML}{FD8D3C}
   \definecolor{cGreen}{HTML}{238B45}
   \definecolor{cRed1}{HTML}{CB181D}
   \definecolor{cRed2}{HTML}{8B1081}
   \definecolor{cYellow}{HTML}{FFC66D}
   \definecolor{cGray}{HTML}{525252}
   \definecolor{cRed1}{HTML}{FF9999}
   \definecolor{cBlue}{HTML}{9999FF}
   \definecolor{cGreen}{HTML}{99FF99}
   
   \newcommand\HiLi[3]{
     \usetikzlibrary{fit,calc}
     \tikz[remember picture, overlay]{
       \node[yshift=1pt,xshift=1pt] (A) {};
       \node[right of=A,yshift=-(#2-1)*\baselineskip,xshift=#3em] (B) {};
       \node[draw=black,rounded corners=3pt,fill=#1,fill opacity=0.5,fit={(A.north)(B.center)}] {};
     }\hspace{-2pt}}
   
   \begin{algorithm}
     \caption{This listing shows batch processing for the \Dfdd~implementation. This
       process starts with a number of initialization steps on the host (highlighted in
       blue). Next, DMs are processed in batches ($idm_{outer}$). For every batch of
       DMs, frequency channels are also processed in batches. This enables overlapping
       of data transfers (highlighted in red) with computation on the GPU
       (highlighted in green). After dedispersion has finished, the output is
       again processed in batches such that postprocessing is overlapped with data
       transfers.}
     \label{alg:fdd}
     \small
     \HiLi{cBlue}{4}{11}
     generate\_spin\_frequency\_table()\;
     allocate\_gpu\_memory()\;
     initialize\_gpu\_fft()\;
     intialize\_jobs()\;
     \HiLi{cRed1}{1}{5}
     copy\_delay\_table()\;
     \For{$idm\_outer = 0 \dots dm\_jobs.size()$} {	\label{alg:fdd-dm_outer}
       \For{$ichannel = 0 \dots channel\_jobs.size()$} {	\label{alg:fdd-channel}
         \tcp{input}
	 \If{$ichannel == 0$} {
	   \HiLi{cRed1}{1}{7.5}
           copy\_to\_gpu($input_{ichannel}$)\;
         }
         \tcp{preprocessing}
	 \HiLi{cGreen}{3}{11.5}
         transpose\_and\_unpack($input_{ichannel}$)\;
	 zero\_pad($input_{ichannel}$)\;
         apply\_r2c\_FFT($input_{ichannel}$)\;
         \tcp{dedispersion}
	 \For{$idm\_inner = 0 \dots ndm\_\buffers$} {	\label{alg:fdd-dm_inner1}
	   \HiLi{cGreen}{1}{16}
           apply\_dedispersion($input_{ichannel}$, $output_{idm\_inner}$)\;
	 } 
         \tcp{input}
         \If{$ichannel + 1 < channel\_jobs.size()$} {
           \HiLi{cRed1}{1}{8}
           copy\_to\_gpu($input_{ichannel+1}$)\;
         }
       } 
       \For{$idm\_inner = 0 \dots ndm\_\buffers$} {	\label{alg:fdd-dm_inner2}
         \tcp{postprocessing}
	 \HiLi{cGreen}{2}{10}
         apply\_c2r\_FFT($output_{idm\_inner}$)\;
         apply\_scaling($output_{idm\_inner}$)\;
         \tcp{output}
         \HiLi{cRed1}{1}{10}
	 copy\_from\_gpu($output_{idm\_inner}$)\;
       }
     } 
   \end{algorithm}

   \section{Results}\label{sec:4}
   \subsection{Performance comparison}
   To allow benchmarking of performance we have generated datasets
   containing simulated astrophysical signals. The datasets cover a
   range of trial-DMs that number from $N_\mathrm{DM}=128$ to
   4096. Each dataset consists of 5 minutes of data with a sampling
   resolution of 64\,$\upmu$s, 1024 channels, 400\,MHz bandwidth and a
   maximum frequency of 1581 MHz, similar to the observational setup
   used by the ongoing \textsc{SUPERB} survey \citep{kbj+18} which
   uses the \textsc{dedisp} library. Trial-DMs start at
   0\,pc\,cm$^{-3}$ and increment by 2\,pc\,cm$^{-3}$ up to the
   maximum specified DM. These datasets are sufficiently large to
   neglect any inaccuracies in CPU or GPU timers. Furthermore, for
   \Dtdd~and \Dfdd~the size of the dataset requires splitting it into
   smaller batches where input and output for one batch is overlapped
   with the computation of another batch.  Measurements were performed
   on a GPU node of the DAS-5 cluster \citep{Bal_2016} and the
   hardware and software setup of the GPU node are listed in
   Table~\ref{tab:setup}.

   \begin{table}
     \caption{The setup used for evaluation of the different implementations}
     \label{tab:setup}
     \centering
     \resizebox{\columnwidth}{!}{
       \begin{tabular}{r | l}
	 CPU & Intel Xeon E5-2660 v3 (one socket, 10 cores/socket) \\
	 GPU & NVIDIA TITAN RTX 24 GiB GDDR6 (PCIe Gen3 16 lanes) \\
	 RAM & 128 GiB DDR4 2133 MHz \\
	 OS &  CentOS 7 kernel 3.10.0-693 (64 bit) \\
	 Compiler & GNU GCC 8.3.0 \\
	 GPU driver & 450.36.06 \\
	 CUDA & CUDA version 11.0.1 kernels compiled for CC 7.5 \\
     \end{tabular}}
   \end{table}

   \begin{figure}[h]
     \centering \includegraphics[width=\columnwidth]{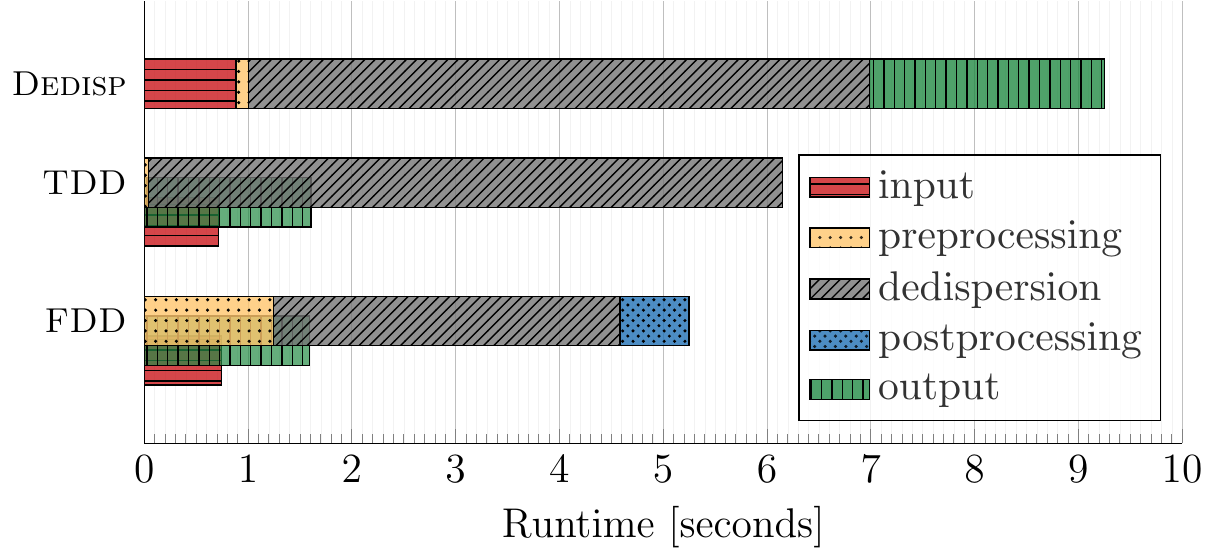} \caption{Runtime
       distribution for \Ddedisp, \Dtdd~and \Dfdd~for 1024
       trial-DMs. Postprocessing for \Dfdd~is optional and depends on the
       processing pipeline.}  \label{fig:runtime-distribution}
   \end{figure}

   \begin{figure}[h]
     \centering \includegraphics[width=\columnwidth]{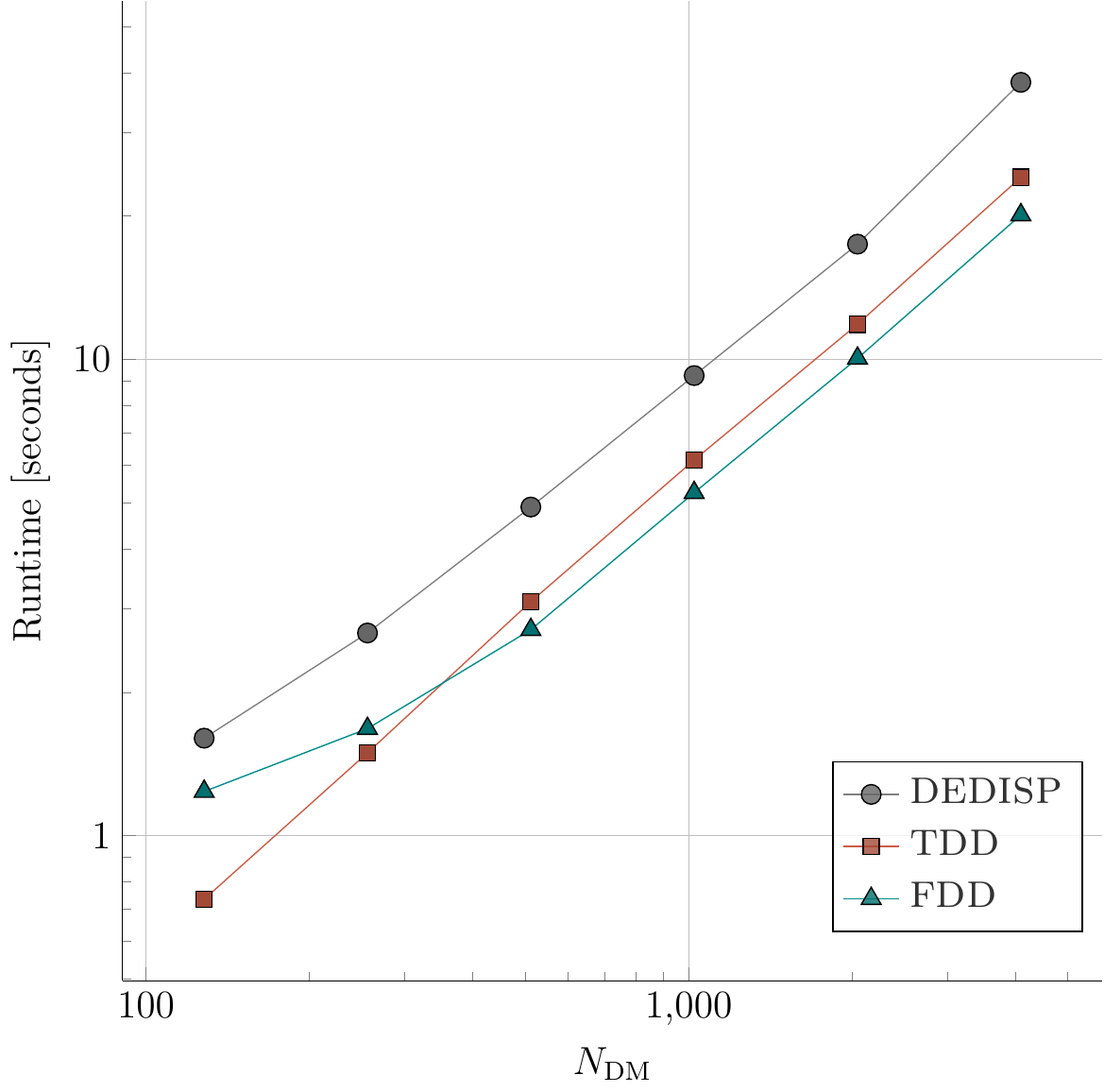} \caption{Comparison
       of runtime for \Ddedisp, \Dtdd~and \Dfdd~for a range of
       trial-DMs that number from 128 to 4096. The runtime
       for \Dfdd~includes postprocessing to transform the output data
       to the time-domain.}  \label{fig:runtime-comparison}
   \end{figure}
   
   Dedispersion is usually a step in a processing pipeline, but for
   this work we have isolated the dedispersion implementation. We
   define the \textit{runtime} of an implementation as the recurring
   time of the implementation which is taken as the summation of
   individual timings of components in the critical path.  The
   critical path for the runtime differs per implementation.

   In Fig.\,\ref{fig:runtime-distribution} we show the distribution of
   runtimes for \Ddedisp, \Dtdd~and \Dfdd.  We find that in
   \Ddedisp~all components (input, preprocessing, dedispersion and
   output) are executed serially. The time required for input and
   output of data is left out of scope in the work by
   \citet{Barsdell_2012} and \citet{Sclocco_2016}. We however argue
   that the time spend on I/O \emph{cannot} be ignored for \Ddedisp,
   as the GPU is idle while the memory transfers between host and
   device take place.

   Our implementation of \Dtdd,~on the other hand, overlaps
   computation on the GPU with I/O which is reflected by a lower
   over-all runtime of a factor 1.5 to 2, depending on the amount of
   trial-DMs. Furthermore, the preprocessing time (transpose and
   unpack) has been reduced while the timing for the dedispersion
   kernel itself is almost the same as in \Ddedisp.

   In \Dfdd,~the input and output are overlapped as well. The
   dedispersion time has been reduced significantly, but at the cost
   of longer preprocessing time (transpose, unpack and FFT) and
   additional postprocessing time (FFT and scaling). We note that most
   of the preprocessing time is spent performing the FFT.  We observe
   a similar distribution of runtime for other numbers of
   trial-DMs. Figure \ref{fig:runtime-comparison} shows a comparison
   of runtime for a range of trial-DMs from 128 to 4096. For a low
   number of trial-DMs \Dtdd~is faster than \Dfdd. Looking at the
   runtime distribution for 128 to 512 DMs we observe that the
   preprocessing time is almost equal from 128 until 512 DMs and then
   increases linearly with the amount of DMs. From 512 DMs onward the
   runtime for \Dfdd~is up to 20\% faster than \Dtdd. The performance
   for a low number of trial-DMs might be improved with tuning of
   batch size parameters. However performance is mostly of interest
   for larger numbers of trial-DMs.  Furthermore, when profiling the
   GPU implementations we observe that, despite part of the runtime of
   the \Dfdd~implementation being limited by the available GPU
   memory\footnote{This is a limitation of the memory size, not to be
   confused with a memory bandwidth limitation. A larger memory size
   would allow for larger and consequently more efficient batch
   sizes.}, the dedispersion algorithm itself is compute bound, except
   for the FFT.

   \subsection{Energy to completion}
   \begin{figure}[h]
     \centering
     \includegraphics[width=\columnwidth]{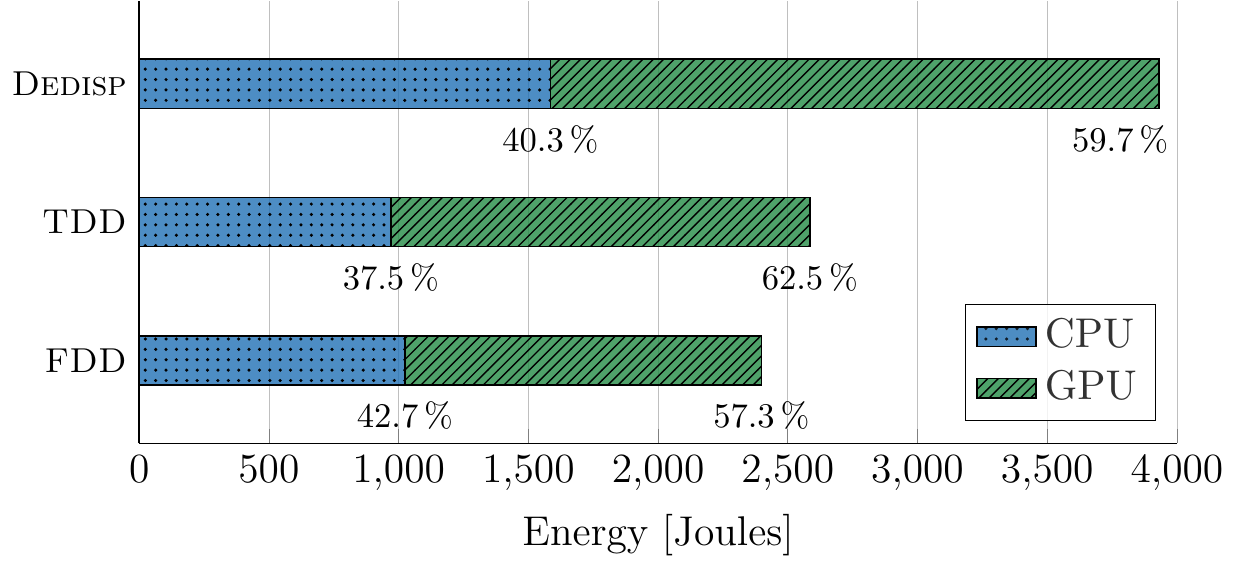}
     \caption{Energy utilization per implementation for 1024 trial-DMs, measured as energy to completion. Percentages shown are the percentage of either the CPU or GPU energy utilization compared to the total per implementation.} \label{fig:powermeasurements}
   \end{figure}

   In many applications the total energy consumption of the system is
   also an important factor. A reduction of runtime of the
   implementation might lead to a reduction in energy as well, but
   does not necessarily scale linearly as compute utilization of the
   device and I/O operations also contribute to the total energy
   utilization. We have measured the energy to completion for our
   three implementations with the same setup used for the runtime
   measurements. Energy usage on the CPU was measured with LIKWID
   \citep{gruber_thomas_2020_4282696} and on the GPU with NVIDIA
   NVML. Figure \ref{fig:powermeasurements} shows the obtained energy
   measurements per implementation. We find that, as one might expect,
   the energy to completion was reduced significantly for both CPU and
   GPU between \Ddedisp~and \Dtdd. The implementation of \Dtdd~leads
   to the same result as \Ddedisp~at only 66\% of the energy
   utilization. Also the \Dfdd~implementation leads to a reduction in
   energy to completion by another 5\%. We observe that the GPU energy
   utilization was reduced compared to \Dtdd, at a slightly increased
   CPU energy utilization.

   \section{Discussion and conclusions}\label{sec:5}
   We have presented the concept of Fourier-domain dedispersion and
   shown that this direct incoherent dedispersion algorithm is a
   viable alternative to the traditional time-domain dedispersion. We
   have implemented the Fourier-domain dedispersion algorithm (\Dfdd)
   in the \Ddedisp\ library by \citet{Barsdell_2012}. An improved
   version of the \Ddedisp\ time-domain dedispersion algorithm (\Dtdd)
   was also added to allow for a fair comparison.

   We find that for a small number of DMs, the time-domain algorithm
   (\Dtdd) is faster than Fourier-domain dedispersion
   (\Dfdd). However, for typical use cases with a large number
   (hundreds) of DMs, \Dfdd\ is the fastest dedispersion
   implementation as it outperforms \Dtdd\ by about 20\%. We also show
   that the energy to completion is lower for \Dtdd\ and \Dfdd\ than
   for \Ddedisp. The energy to completion for 1024 trial-DMs with
   \Dfdd\ is only 61\% of the energy utilization of \Ddedisp~on the
   system used. Furthermore, where \Ddedisp\ is largely I/O bound
   \citep{Sclocco_2016}, we were able to alleviate some of the I/O
   bottlenecks in \Dtdd: we optimized the memory transfers, and we
   re-implemented the kernels that pre-process input data. While some
   parts of \Dfdd\ (such as the FFTs) are also I/O bound, the most
   dominant \Dfdd\ dedispersion kernel is compute bound.  We therefore
   expect \Dfdd\ to scale better with advances in GPU technology,
   where compute performance typically increases much faster than
   memory bandwidth.

   For pulsar surveys using FFT-based periodicity searches, the
   \Dfdd\ algorithm provides even higher performance compared to the
   \Ddedisp\ and \Dtdd\ algorithms. For the time-domain dedispersion
   algorithms, the FFT-based periodicity searches would perform an FFT
   of the dedispersed time-series to produce power spectra to search
   of periodic signals. By using the \Dfdd\ algorithm this FFT of the
   periodicity search, as well as the \Dfdd\ postprocessing step in
   which the data is Fourier-transformed back to the time-domain can
   be omitted. Instead, the amplitudes and phases of the dedispersed
   spectra from \Dfdd\ can be directly used for the subsequent
   processing steps of periodicity searches, like Fourier-domain
   acceleration searches \citep{ran01}. In any situation where the
   number of DMs exceeds the number of (observing) frequency channels,
   Fourier-domain dedispersion will require fewer FFTs for
   dedispersion compared to time-domain dedispersion.

   Besides the algorithmic optimizations proposed in
   \ref{ssec:alg_optimisation}, we consider some optimizations that
   may further improve performance. First, the throughput of the
   current implementation of Fourier-domain dedispersion depends on
   parameters such as the data size (e.g.\ number of channels, number
   of DMs), as well as the amount of available GPU memory and the PCIe
   bandwidth. Autotuning \citep{Sclocco_2016} may be useful to find
   the best-performing set of tuning parameters, such as the optimal
   batch dimensions.

   Second, the current implementation performs all operations in
   32-bit floating point.  The use of 16-bit half precision or even
   8-bit integers would not only double/quadruple the amount of
   samples that can be processed in one batch, it would also allow the
   use of \emph{tensor cores\/} to perform the dedispersion (tensor
   cores are special-function units in modern GPUs, that perform
   mixed-precision matrix multiplications much faster than regular GPU
   cores).  Unfortunately, neither 8-bit FFTs, nor tensor-core
   dedispersion is simple to implement. In another signal-processing
   application (a correlator), tensor cores are 5 to 10 times more
   (energy) efficient compared to regular GPU cores \citep{rom21}.

   Furthermore, the performance of the Fourier-domain algorithm can be
   improved by reusing phase rotations (Eqn.\,\ref{eq:phasors}) when
   multiple observations with the same observational setup are
   available. This is the case for telescope where signals from
   multiple beams are obtained simultaneously, or for survey
   observations with many similar observations. We note that this
   approach reduces the number of DMs or time samples that can be
   processed in one batch, as the amount of available GPU memory is
   limited, and requires careful tuning.

   \begin{acknowledgements}
     We thank Scott Ransom for very constructive comments on the
     manuscript and are indebted to Andrew Jameson for help
     incorporating \textsc{fdd} into the \textsc{dedisp} library.
     This project received funding from the Netherlands Organization
     for Scientific Research (NWO) through the Fourier-Domain
     Dedispersion (OCENW.XS.083) and the DAS-5 (621.018.201) grants.
   \end{acknowledgements}
   
   \bibliographystyle{aa}

\begin{thebibliography}{28}
\expandafter\ifx\csname natexlab\endcsname\relax\def\natexlab#1{#1}\fi

\bibitem[{Bal {et~al.}(2016)}]{Bal_2016}
Bal, H. {et~al.} 2016, IEEE Computer, 49, 54

\bibitem[{Barsdell {et~al.}(2012)Barsdell, Bailes, Barnes, \&
  Fluke}]{Barsdell_2012}
Barsdell, B.~R., Bailes, M., Barnes, D.~G., \& Fluke, C.~J. 2012, MNRAS, 422,
  379–392

\bibitem[{{Bassa} {et~al.}(2017{\natexlab{a}}){Bassa}, {Pleunis}, \&
  {Hessels}}]{bph17}
{Bassa}, C.~G., {Pleunis}, Z., \& {Hessels}, J.~W.~T. 2017{\natexlab{a}},
  Astronomy and Computing, 18, 40

\bibitem[{{Bassa} {et~al.}(2017{\natexlab{b}}){Bassa}, {Pleunis}, {Hessels},
  {Ferrara}, {Breton}, {Gusinskaia}, {Kondratiev}, {Sanidas}, {Nieder},
  {Clark}, {Li}, {van Amesfoort}, {Burnett}, {Camilo}, {Michelson}, {Ransom},
  {Ray}, \& {Wood}}]{bph+17}
{Bassa}, C.~G., {Pleunis}, Z., {Hessels}, J.~W.~T., {et~al.}
  2017{\natexlab{b}}, \apjl, 846, L20

\bibitem[{{Bhattacharyya} {et~al.}(2016){Bhattacharyya}, {Cooper}, {Malenta},
  {Roy}, {Chengalur}, {Keith}, {Kudale}, {McLaughlin}, {Ransom}, {Ray}, \&
  {Stappers}}]{bcm+16}
{Bhattacharyya}, B., {Cooper}, S., {Malenta}, M., {et~al.} 2016, \apj, 817

\bibitem[{{Bracewell}(1986)}]{Bracewell_1986}
{Bracewell}, R.~N. 1986, {The Fourier Transform and its applications}

\bibitem[{Gruber {et~al.}(2021)Gruber, Eitzinger, Hager, \&
  Wellein}]{gruber_thomas_2020_4282696}
Gruber, T., Eitzinger, J., Hager, G., \& Wellein, G. 2021, RRZE-HPC/likwid:
  likwid-4.2.1

\bibitem[{{Hankins}(1971)}]{han71}
{Hankins}, T.~H. 1971, \apj, 169, 487

\bibitem[{{Hankins} \& {Rickett}(1975)}]{hr75}
{Hankins}, T.~H. \& {Rickett}, B.~J. 1975, Methods in Computational Physics,
  14, 55

\bibitem[{Jameson \& Barsdell(2019)}]{Heimdall}
Jameson, A. \& Barsdell, B.~R. 2019, Heimdall Transient Detection Pipeline

\bibitem[{{Kaspi} \& {Kramer}(2016)}]{kk16}
{Kaspi}, V.~M. \& {Kramer}, M. 2016, arXiv e-prints, arXiv:1602.07738

\bibitem[{{Keane} {et~al.}(2018){Keane}, {Barr}, {Jameson}, {Morello}, {Caleb},
  {Bhandari}, {Petroff}, {Possenti}, {Burgay}, {Tiburzi}, {Bailes}, {Bhat},
  {Burke-Spolaor}, {Eatough}, {Flynn}, {Jankowski}, {Johnston}, {Kramer},
  {Levin}, {Ng}, {van Straten}, \& {Krishnan}}]{kbj+18}
{Keane}, E.~F., {Barr}, E.~D., {Jameson}, A., {et~al.} 2018, \mnras, 473, 116

\bibitem[{{Keith} {et~al.}(2010){Keith}, {Jameson}, {van Straten}, {Bailes},
  {Johnston}, {Kramer}, {Possenti}, {Bates}, {Bhat}, {Burgay}, {Burke-Spolaor},
  {D'Amico}, {Levin}, {McMahon}, {Milia}, \& {Stappers}}]{kjs+10}
{Keith}, M.~J., {Jameson}, A., {van Straten}, W., {et~al.} 2010, \mnras, 409,
  619

\bibitem[{{Kulkarni}(2020)}]{kul20}
{Kulkarni}, S.~R. 2020, arXiv e-prints, arXiv:2007.02886

\bibitem[{{Lazarus} {et~al.}(2015){Lazarus}, {Brazier}, {Hessels},
  {Karako-Argaman}, {Kaspi}, {Lynch}, {Madsen}, {Patel}, {Ransom}, {Scholz},
  {Swiggum}, {Zhu}, {Allen}, {Bogdanov}, {Camilo}, {Cardoso}, {Chatterjee},
  {Cordes}, {Crawford}, {Deneva}, {Ferdman}, {Freire}, {Jenet}, {Knispel},
  {Lee}, {van Leeuwen}, {Lorimer}, {Lyne}, {McLaughlin}, {Siemens}, {Spitler},
  {Stairs}, {Stovall}, \& {Venkataraman}}]{lbh+15}
{Lazarus}, P., {Brazier}, A., {Hessels}, J.~W.~T., {et~al.} 2015, \apj, 812, 81

\bibitem[{{Lorimer} \& {Kramer}(2012)}]{lk12}
{Lorimer}, D.~R. \& {Kramer}, M. 2012, {Handbook of Pulsar Astronomy}
  ({Cambridge University Press})

\bibitem[{{Lyon} {et~al.}(2016){Lyon}, {Stappers}, {Cooper}, {Brooke}, \&
  {Knowles}}]{lbc+16}
{Lyon}, R.~J., {Stappers}, B.~W., {Cooper}, S., {Brooke}, J.~M., \& {Knowles},
  J.~D. 2016, \mnras, 459, 1104

\bibitem[{{Manchester} {et~al.}(2001){Manchester}, {Lyne}, {Camilo}, {Bell},
  {Kaspi}, {D'Amico}, {McKay}, {Crawford}, {Stairs}, {Possenti}, {Kramer}, \&
  {Sheppard}}]{mlc+01}
{Manchester}, R.~N., {Lyne}, A.~G., {Camilo}, F., {et~al.} 2001, \mnras, 328,
  17

\bibitem[{{Manchester} \& {Taylor}(1972)}]{mt72}
{Manchester}, R.~N. \& {Taylor}, J.~H. 1972, \aplett, 10, 67

\bibitem[{{Petroff} {et~al.}(2019){Petroff}, {Hessels}, \& {Lorimer}}]{phl19}
{Petroff}, E., {Hessels}, J.~W.~T., \& {Lorimer}, D.~R. 2019, \aapr, 27, 4

\bibitem[{{Pleunis} {et~al.}(2017){Pleunis}, {Bassa}, {Hessels}, {Kondratiev},
  {Camilo}, {Cognard}, {Grie{\ss}meier}, {Stappers}, {van Amesfoort}, \&
  {Sanidas}}]{pbh+17}
{Pleunis}, Z., {Bassa}, C.~G., {Hessels}, J.~W.~T., {et~al.} 2017, \apjl, 846,
  L19

\bibitem[{{Ransom}(2001)}]{ran01}
{Ransom}, S.~M. 2001, PhD thesis, Harvard University

\bibitem[{Romein(2021)}]{rom21}
Romein, J.~W. 2021, A\&A, to appear

\bibitem[{Sclocco {et~al.}(2016)Sclocco, van Leeuwen, Bal, \& van
  Nieuwpoort}]{Sclocco_2016}
Sclocco, A., van Leeuwen, J., Bal, H., \& van Nieuwpoort, R. 2016, Astronomy
  and Computing, 14

\bibitem[{{Sclocco} {et~al.}(2015){Sclocco}, {van Leeuwen}, {Bal}, \& {van
  Nieuwpoort}}]{Sclocco_2015}
{Sclocco}, A., {van Leeuwen}, J., {Bal}, H.~E., \& {van Nieuwpoort}, R.~V.
  2015, in 2015 IEEE Global Conference on Signal and Information Processing
  (GlobalSIP), 468--472

\bibitem[{{Stovall} {et~al.}(2014){Stovall}, {Lynch}, {Ransom}, {Archibald},
  {Banaszak}, {Biwer}, {Boyles}, {Dartez}, {Day}, {Ford}, {Flanigan}, {Garcia},
  {Hessels}, {Hinojosa}, {Jenet}, {Kaplan}, {Karako-Argaman}, {Kaspi},
  {Kondratiev}, {Leake}, {Lorimer}, {Lunsford}, {Martinez}, {Mata},
  {McLaughlin}, {Roberts}, {Rohr}, {Siemens}, {Stairs}, {van Leeuwen},
  {Walker}, \& {Wells}}]{slr+14}
{Stovall}, K., {Lynch}, R.~S., {Ransom}, S.~M., {et~al.} 2014, \apj, 791, 67

\bibitem[{{Taylor}(1974)}]{tay74}
{Taylor}, J.~H. 1974, \aaps, 15, 367

\bibitem[{{Zackay} \& {Ofek}(2017)}]{zo17}
{Zackay}, B. \& {Ofek}, E.~O. 2017, \apj, 835, 11

\end{thebibliography}

\end{document}